\documentclass[12pt,preprint]{aastex}
\shortauthors{Li J. Z. et al.}
\shorttitle{Infant star taking a UV shower}

\begin{document}

\title{The Discovery of a Disk-Jet System Directly Exposed to Strong UV Fields in the Rosette Nebula}
\author{Jin Zeng Li$^{1}$, Travis A. Rector$^{2}$}

\affil{$^{1}$National Astronomical Observatories, Chinese Academy of 
           Sciences, Beijing 100012, China (E-mail: ljz@bao.ac.cn) \\
$^{2}$University of Alaska Anchorage, Anchorage, AK  99508, USA \\
}

\begin{abstract} 

We report on the discovery of an optical jet with a striking morphology
in the Rosette Nebula.  It could be the most extreme case known
of an accretion disk and jet system directly exposed to strong ionization
fields that impose strong effects on its disk evolution.
Unlike typical optical flows, this jet system is found to have a high
excitation nature mainly due to disruptive interaction with the violent
environment.  As a result, the extension of the highly-collimated jet and 
possible former episodes of the degenerated counterjet all show bow-shocked 
structures.
%  Optical spectroscopy of the apparent power source, which 
%suffers from very little extinction rather than being shrouded in an 
%opaque extended envelope, reveals signatures of supersonic mass inflow.
%However, there's no simultaneous detection of noticable veiling or blue excess,
%which would signify mass accretion onto the central young stellar object. 
%CaII emission, as indicators of enhanced chromospheric activity, are also
%absent.  
Our results provide implications on how incipience of massive 
stars in giant molecular clouds prevents further generations of low-mass
star formation, and possibly also how isolated substellar/planetary mass
objects in regions of massive star formation are formed.

\end{abstract}


\section{Introduction}

 Herbig-Haro (HH) jets are the optical manifestations of
collimated outflows from young stellar objects (YSOs) in their early
stages of evolution and, along with typical HH objects, are apparent
tracers of active star formation. YSOs, unfortunately, are usually deeply
embedded in dense, extended envelopes and/or opaque molecular cloud
cores that strongly impede efforts for investigations in the optical 
regime and make our view of the early stages of star formation still 
quite a puzzle. In some cases, however, YSOs are located within
the confines of HII regions (Reipurth et al. 1998) and their power sources
suffer very low extinction and are therefore optically visible.  Similar
jets immersed in a photoionized medium were identified in the Orion
nebula and near the late B stars in NGC~1333 (Bally et al. 2000;
Bally \& Reipurth 2001).
However, some properties of photoionized jets in HII regions are
found to be highly dependent on their environment.  Those immersed in the 
outskirts of HII regions, or near soft-UV radiation sources such as late B stars,
tend to emerge as bipolar jets which usually share a ``C" shaped appearance
(Bally et al. 2000; Bally \& Reipurth 2001), whereas jets deeply embedded 
in HII regions 
endure a more disruptive environment and seem to favor highly asymmetric
or even monopolar jet formation (Reipurth et al. 1998).

The Rosette Nebula is a spectacular HII region excavated by the strong
stellar winds from O stars in the center of the young open
cluster NGC~2244, which has an age of about 3 x $10^6$ yr (Ogura \& Ishida 1981).
This on-going star forming region (Meaburn \& Walsh 1986;
Clayton et al. 1998; Phelps \& Lada 1997; Li et al. 2002; Li 2003) is located
at a distance of $\sim$1500 pc at the tip of a giant molecular cloud
complex with an extent of around 100 pc (Dorland \& Montmerle 1987).
In this paper, we present the discovery of an optical jet system embedded 
in the Rosette Nebula with a fascinating morphology that displays exceptional 
new features not known before.

\section{Observations and Data Reduction}

\subsection{Narrowband Imaging}

Narrow band images of the Rosette Nebula were obtained 3 March 1999 with the
Kitt Peak National Observatory 0.9-meter telescope and the 8k$\times$8k MOSAIC I
camera (Muller et al. 1998). Images were taken with the H${\alpha}$,
[OIII] and [SII] filters, whose central wavelengths/FWHMs are 6569\AA/80\AA, 5021\AA/55\AA,
and 6730\AA/80\AA, respectively.  For each filter, five exposures of 600~seconds
were taken, each image slightly offset to fill in physical gaps between the
MOSAIC CCDs.  The pixel scale is 0.423\arcsec~pixel$^{-1}$, resulting in a 59\arcmin~x~59\arcmin\
field of view. An astrometric solution for these images was determined with
the use of stars in the Guide Star Catalog 1.2 (Roser et al. 1998). The astrometric
accuracy in the optical images presented is $\sim$1\arcsec.

\subsection{Optical Spectroscopy}

Optical spectroscopy of the jet system, with a
200\AA~mm$^{-1}$, 4.8\AA~pixel$^{-1}$ dispersion, and a 2.5\arcsec slit,
was carried out with the 2.16m telescope of the National Astronomical
Observatories of China on 10 and 16 January 2003, with slit positions
along and orthogonal to the jet direction. A possible
physical companion of the apparent power source was also observed with the
orthogonal slit position to avoid contamination from the jet system.
An OMR (Optomechanics Research Inc.) spectrograph and a Tektronix 1024${\times}$1024 CCD
detector were used in this run of observations.

The spectral data were reduced following
standard procedures in the NOAO Image Reduction and Analysis Facility
(IRAF, version 2.11) software package. The CCD reductions included bias
and flat-field correction, successful nebular background subtraction, and
cosmic rays removal. Wavelength calibration was performed based on helium-argon
lamps exposed at both the beginning and the end of the observations every night.
Flux density calibration of each spectrum was conducted based on observations of at
least two of the KPNO spectral standards (Massey et al. 1988) per night.

\section{Results and Discussion}

\subsection{The Jet System}

An optical jet system with a striking morphology was identified
when scrutinizing the narrow band images of the Rosette Nebula (Fig. 1). A
prominent jet with a position angle of 312\arcdeg\ is clearly seen traced back
to a faint visible star, indicating rather low extinction along the
line of sight. The circumstellar envelope of the YSO
might have been stripped away, leaving a photoablating disk exposed
to the strong UV radiation field of the exciting sources in the HII region.
The optical jet remains highly collimated for a projected distance
of $>$8000~AU  from the energy source if a mean distance
of 1500 pc to Rosette is adopted.  At the end of the collimated jet is a 
side-impacted shock structure, probably due to severe radiation pressure from the
strong UV field present and strong stellar winds encountered.   
A prominent knot or compact mass bullet resembling a point source is immediately
noticed in the highly collimated part of the jet with a projected separation of
7\arcsec\ from the source.  Another one or 
two knots can be marginally noticed in the bow shaped extension of the jet, 
indicating episodic eruption events from the driving source.
% which could well be an example case of ablated jet from a low mass YSO ?

 We infer the possible existence of 
a counterjet from the presence of a promising bow shock
structure on the opposite side of the jet.  Successive episodes
of counterjet ejection may have ceased or may be too faint. 
Alternatively, this structure may simply be shocked
interstellar material that was swept up and photoionized by the strong
radiation field, which happen to be projected onto the vicinity of
the jet exciting source. Another possible explanation is that it is
disrupted jet material from a possible T Tauri companion (please refer to
Fig.~1) or from nearby young stars. If it is indeed a 
degenerated counterjet, however, it may be the only existing observational 
evidence for how bipolar jets evolve into monopolar or 
highly asymmetric jets.
We further argue that the long-puzzling high-velocity components from
the photoionized gas fronts aggregated in the Rosette (Meaburn \& Walsh 1986; Clayton \& Meaburn 1995; Clayton, Meaburn \& Lopez et al. 1998) could
actually be from rapidly disrupted young disk-jet systems in the inner
part of the HII region.  Diffuse [OIII] emission is prominent from 
all parts of the jet system, which is likely due to its photo-dissipating nature.  
[SII] emission from the highly-collimated part
of the jet decreases rapidly from the base of the jet and is undetected
in the shocked structures, indicating a steep decline of internal
shock effects or rather overwhelming external ionization.

Spectroscopic observation along the jet direction reveals prominent H$\alpha$ 
($W_{\lambda} \sim$~35\AA) as well as [SII] emission.  The presence of
significant [OIII] $\lambda\lambda$4959, 5007 line emission 
confirms our suspicion that the jet is in a high-excitation state.
Fig. 2 clearly shows the existence of a weak counterjet as discussed above.
Note that enhanced [NII] emission is detected along the jet, which is not
shown here.

\subsection{The exciting source}

Spectroscopic observations of the exciting source reveal primarily the 
spectrum of a normal F8Ve star with weak H$\alpha$ ($W_{\lambda} \sim$~6.7\AA)
and prominent [OIII] emission.  However, the spectrum also shows a 
significant red-displaced absorption component in H$\alpha$, with 
a receding velocity of 500$\pm$50~km~s$^{-1}$ (Fig. 3). Two weak
but significant emission features with unclear nature are also present immediately
to the red of the rest H$\alpha$ emission and its red-displaced absorption. One is
centered at 6582.5 \AA~ with an accuracy of $\pm$ 1 \AA~, and can be 
the pronounced [NII] $\lambda$ 6583.6 emission detected also along the jet. 
This gives a high [NII] to H$\alpha$ flux ratio of about 1:3.
Albeit the other with a central wavelength of 6613.3 \AA~ cannot be identified 
with any known emission lines with rest velocities. A possible interpretation
is a red-displaced emission component of H$\alpha$. This however would suggest
an extraordinarily high receding velocity of 2300$\pm$50~km~s$^{-1}$ and should
be treated with great caution. The possible ejection of high velocity, compact 
mass bullets from the energy source, if true, can be partially supported by the 
clear existence of a compact knot in the highly collimated part of the jet
as mentioned in the previous subsection, and may offer a big challenge to 
currently available disk-jet models. This scenario, however, seems to match well
with Meaburn's (2003) recent speculations on the formation of ablated jets
in HII regions.
% introduces highly-prejudiced jet driving from the power source in opposite directions.  

  Spectral energy distribution (Kurucz 1991) fitting of the energy source
based on spectroscopic observations and data extracted from 2MASS indicated a relic
disk mass of only 0.006 M$_{\odot}$, consistent with a photoablating
origin of the system. It is noteworthy that Inverse P Cygni (IPC) profiles
seldom appear in low Balmer
emission lines such as H$\alpha$.  The most well known case is the YY Orionis
type, weak-lined T Tauri star T Cha (Alcala, Covino \& Franchini et al. 1993), 
from which, at most, a relic circumstellar
disk can be expected. An IPC profile in H$\alpha$ is usually attributed to its
optically thin nature, which normally leads to the conclusion of a high
inclination angle of the relic disk. Furthermore, this must happen under
tight conditions such as low-density in the residual circumstellar
material and weak H$\alpha$ emission (Alcala, Covino \& Franchini et al. 1993).
Both are consistent with properties of this jet system.

\subsection{Implications}

A similar red-displaced absorption profile is also extracted 
around the power source from the spectra taken along the jet direction, 
which helps to elucidate the reliability of the spectral reduction and time 
elapsed mass accretion of the central YSO.  The simultaneous existence of 
signatures of mass inflow and outflow, along with the absence of apparent 
veiling or blue excess, the absence of signatures of chromospheric 
activity, and its weak or absent Balmer emission in combination is strong evidence 
that the forming star is highly affected by its location in the strong UV 
radiation fields.  Indeed, Reipurth et al. (1998) state that external ionization may 
help on the feeding and/or launching of the jet, which makes jet formation in such
systems survive. We further speculate that the aforementioned effects eventually 
change the configuration of the stellar-disk magnetosphere 
such that photodissipated material in the relic disk is 
easily loaded onto the magnetic channels, especially on the side facing the
strong radiation fields.  It is conceivable that most or all of this material could 
have been ejected in the form of a jet on the opposite side of the 
mass loading, instead of eventually ramming into the
contracting infant star embarrassed with starvation. 
It's therefore hard for the energy source to further grow in mass, its accretion
disk could be dissipated on a comparatively short time scale and its evolution
toward the main sequence is either highly accelerated or in some cases even led 
to the formation of a failed star that would have evolved into a higher mass
star under normal conditions.
Many young, very low mass stars and brown dwarfs 
are found to show evidence of mass accretion from its circumstellar environment 
(Fernandez \& Comeron 2001; Luhman, Briceno \& Stauffer 2003; Jayawardhana, 
Mohanty \& Basri 2003; White \& Basri, 2003).  This study presents
further implications on how incipience of 
massive stars in giant molecular clouds inhibits further generations of low-mass 
star formation.  It could also well serve as an inspiring case of how isolated 
substellar/planetary mass objects in regions of massive star formation, especially 
those found in HII regions (Zapatero Osorio et al. 2000), were prohibited from 
growing in mass, ceased their accretion process and came into being.

{\flushleft \bf Acknowledgments~}

We are grateful to an anonymous referee for the 
many helpful comments on the paper. Thanks prof. You-Hua Chu very much
for her valuable comments and suggestions.  Appreciations 
to prof. W. P. Chen and W. H. Ip for their kind accommodations and help 
during my two years stay at NCU. This work has made use of the 2MASS database.

\bibliographystyle{aa}

\clearpage

\figcaption[f1.eps]{The H$\alpha$ imaging of the jet system (3\arcmin.2 x 2\arcmin.7).  North
is up and east is to the left. The prominent
jet, at the lower center of the image, and its extended bow shock structure as a 
whole extended to $\sim$0.13 pc away from the driving source. Assuming that the 
jet has an inclination of 45\arcdeg\ off the plane of the sky and a typical flow 
velocity of 150~km~s$^{-1}$, a lower
limit of the kinematic age of the jet of $\sim$1200~yr is derived. The apparent 
power source is located
at R.A. = 06h32m20.76s, Dec = 04d53m02.9s (J2000). A possible companion also exists, for which
the coordinates from 2MASS are R.A. = 06h32m20.79s, Dec = 04d52m58.1s (J2000). 
Optical spectroscopy of this star shows prominent Balmer
emission lines ($W_{\lambda}$ of H$\alpha$ and H$\beta$ are 72\AA\ \& 51\AA, respectively)
are superimposed on spectrum of a low-reddening K dwarf. 
Both components of the possible binary system are therefore T Tauri stars 
previously unknown in this region. However, no strong intervention between 
the two is apparent.}

\figcaption[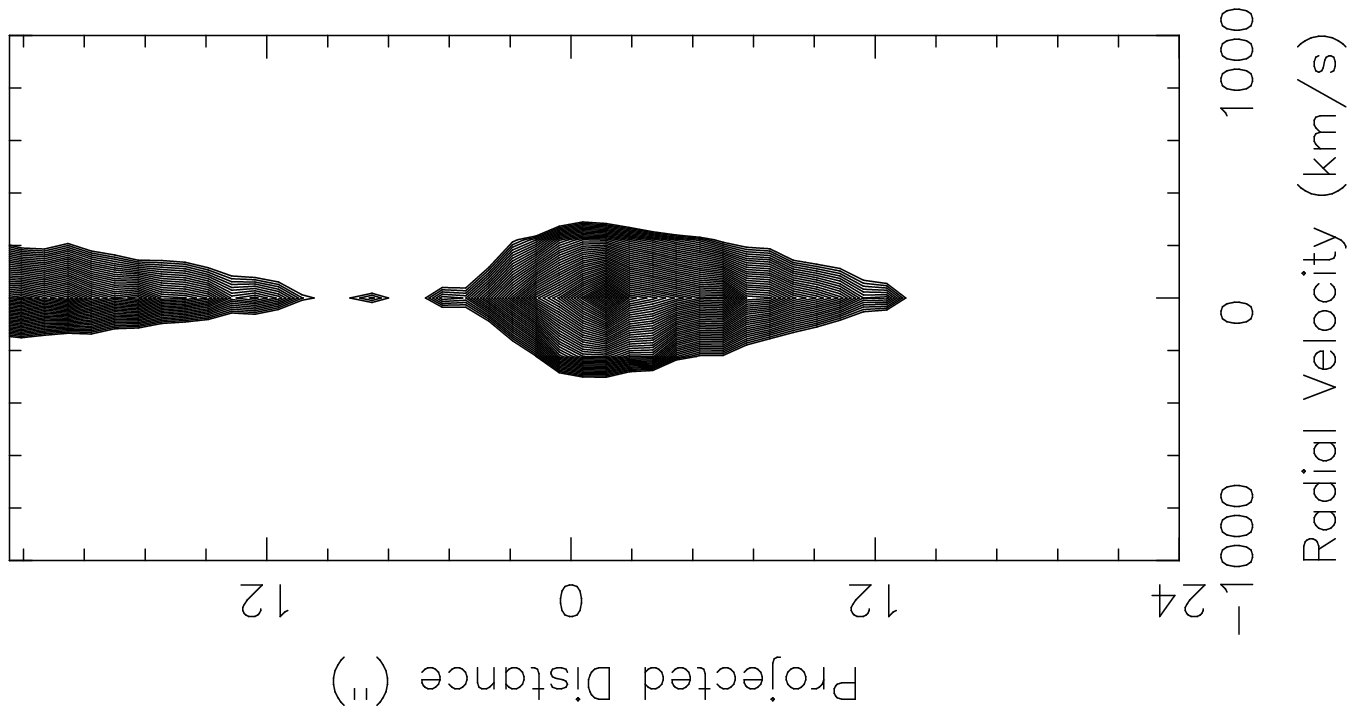]{Contour map of the H$\alpha$ emission along the jet direction.
Note that low-resolution spectroscopy was employed
in this run of observations and no prominent velocity difference between the 
opposite jets can be noticed. However, there is clear indication 
of the existence of a poorly developed or degenerated counterjet.
Compact H$\alpha$ emission at the upper part of the map is from the shocked 
structure located on the opposite side of the prominent jet.}

\figcaption[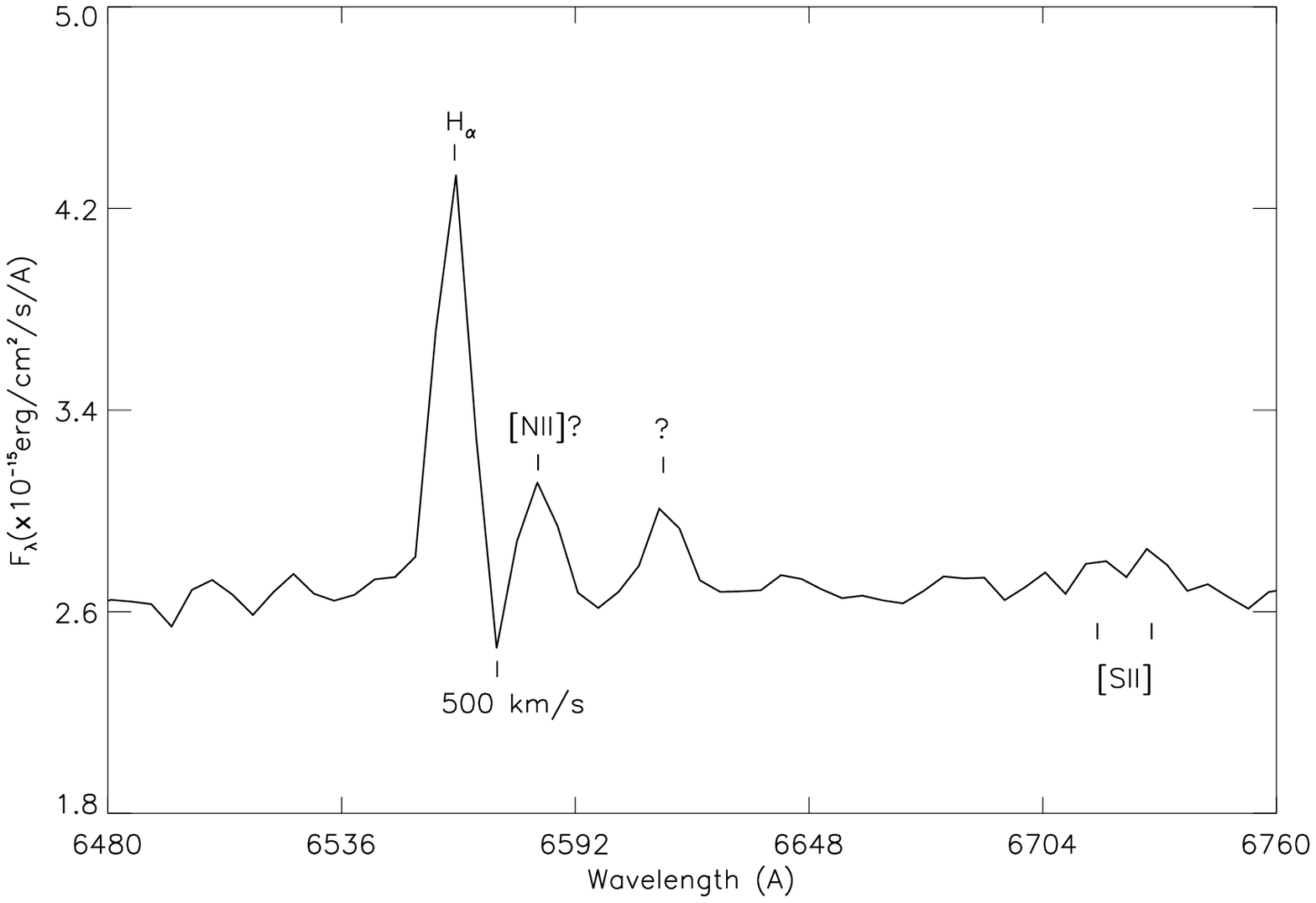]{The H$\alpha$ emission from the jet driving source.
Note the red-displaced absorption component in the stellar continuum emission.
%The inset is a position-velocity plot of the features detected
%around H$\alpha$.  Intense H$\alpha$ emission
%at the upper part of the inset is again from the shocked structure at the
%opposite side of the jet.
}

%%%UCP%%%
\newpage
\plotone{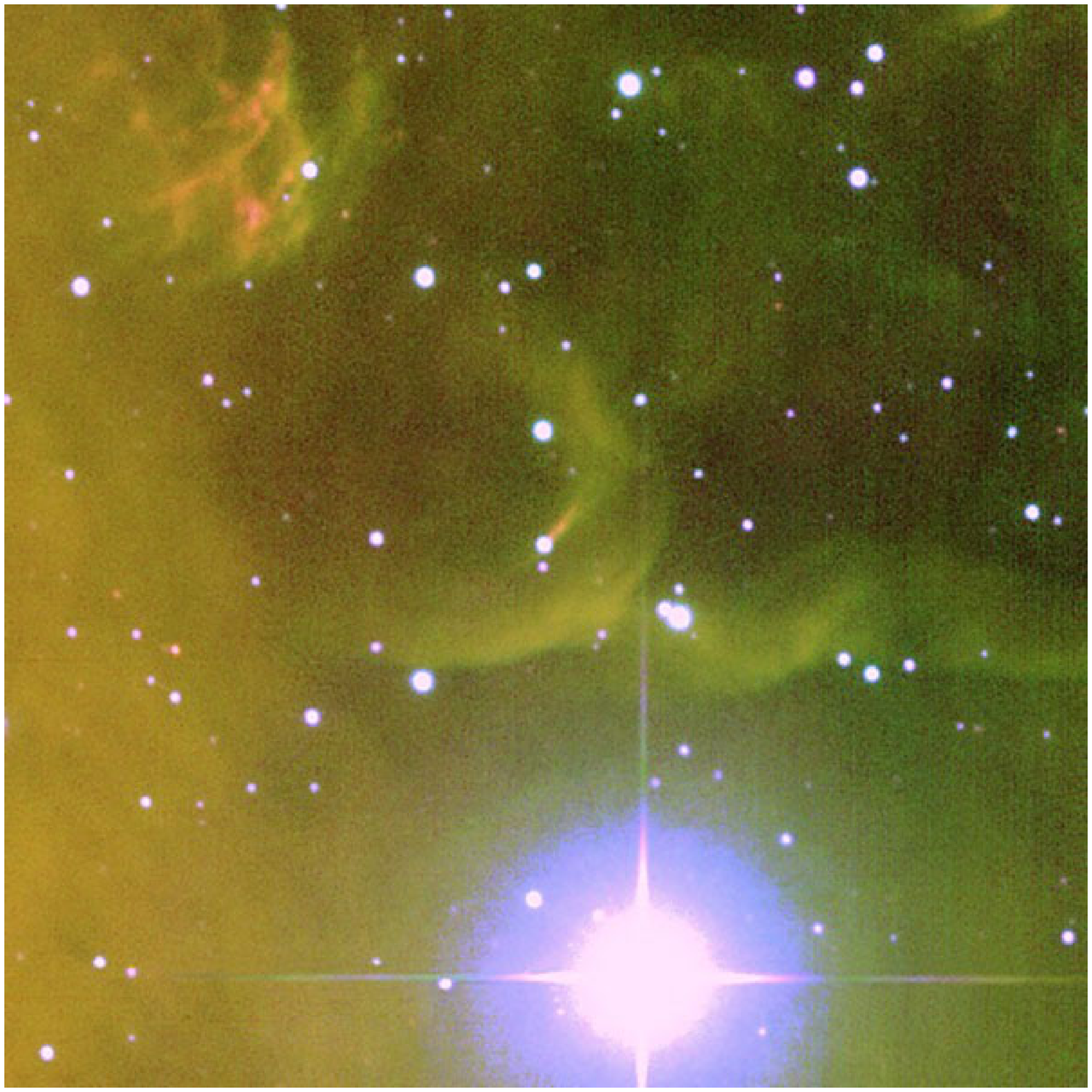}
\newpage
\plotone{f2.eps}
\newpage
\plotone{f3.eps}
%\plotone{f3b.eps}

\end{document}